\documentclass[runningheads]{llncs}
\usepackage{graphicx}
\usepackage{datetime}
\usepackage{lipsum}
\usepackage{fancyhdr}
\usepackage{afterpage}
\usepackage[english]{babel}
\usepackage{amsmath}
\usepackage{tabularx}
\usepackage{multirow}
\usepackage{amssymb}
\usepackage{wrapfig}
\usepackage{svg}
\usepackage{hyperref}

\usepackage{tikz,xcolor,hyperref}

\definecolor{lime}{HTML}{A6CE39}
\DeclareRobustCommand{\orcidicon}{%
	\begin{tikzpicture}
	\draw[lime, fill=lime] (0,0) 
	circle [radius=0.16] 
	node[white] {{\fontfamily{qag}\selectfont \tiny ID}};
	\draw[white, fill=white] (-0.0625,0.095) 
	circle [radius=0.007];
	\end{tikzpicture}
	\hspace{-2mm}
}

\foreach \x in {A, ..., Z}{%
	\expandafter\xdef\csname orcid\x\endcsname{\noexpand\href{https://orcid.org/\csname orcidauthor\x\endcsname}{\noexpand\orcidicon}}
}

\begin{document}

\makeatletter
\def\old@comma{,}
\catcode`\,=13
\def,{%
  \ifmmode%
    \old@comma\discretionary{}{}{}%
  \else%
    \old@comma%
  \fi%
}
\makeatother

\title{Detecting Anomalous Events in Object-centric Business Processes via Graph Neural Networks}
\titlerunning{Detecting Anomalous Events in Object-Centric Business Processes via GNNs}
\author{Alessandro Niro\orcidA{} \and
Michael Werner\orcidB{}}
\authorrunning{A. Niro and M. Werner}
\institute{University of Amsterdam, Amsterdam, The Netherlands \\
\email\{a.niro, m.werner\}@uva.nl}
\maketitle              % typeset the header of the contribution
\begin{abstract}
Detecting anomalies is important for identifying inefficiencies, errors, or fraud in business processes. Traditional process mining approaches focus on analyzing ‘flattened’, sequential, event logs based on a single case notion. However, many real-world process executions exhibit a graph-like structure, where events can be associated with multiple cases. Flattening event logs requires selecting a single case identifier which creates a gap with the real event data and artificially introduces anomalies in the event logs. Object-centric process mining avoids these limitations by allowing events to be related to different cases. This study proposes a novel framework for anomaly detection in business processes that exploits graph neural networks and the enhanced information offered by object-centric process mining. We first reconstruct and represent the process dependencies of the object-centric event logs as attributed graphs and then employ a graph convolutional autoencoder architecture to detect anomalous events. Our results show that our approach provides promising performance in detecting anomalies at the activity type and attributes level, although it struggles to detect anomalies in the temporal order of events.

\end{abstract}

\section{Introduction}
\label{sect:introduction}

Process mining aims to discover, monitor, and enhance existing business processes by leveraging data traces generated by their execution and stored into event logs \cite{vanderAalst2022}. Business processes can be defined as a set of activities that enable the organization to achieve a specified goal. They are seldom friction-less. Errors, inefficiencies, and fraud during process executions can lead to significant losses for the organizations. The ability to detect and mitigate harmful anomalies is crucial for maintaining the effectiveness and efficiency of business operations.

Traditional approaches to process mining rely on `flattened' event logs. Events are characterized by a single case identifier \cite{van_der_aalst_object-centric_2019} and process instances are assumed to be strictly ordered sequences of events. Anomaly detection techniques have primarily centered on approaches applied to flattened event logs. Conformance checking approaches \cite{bezerra_anomaly_2009,bezerra_algorithms_2013,bohmer_multi-perspective_2016} have focused on identifying deviations of sequential process instances, as captured by the event logs, from an \textit{a priori} process model, which is a necessary input for this type of techniques. Machine learning approaches, based on distance measures \cite{junior_anomaly_2020,tavares_analysis_2020} or reconstruction errors \cite{nolle_analyzing_2018,nolle_binet_2022,nguyen_autoencoders_2019,huo_graph_2021}, have focused on detecting anomalies directly from event logs, adopting methods suited for strictly ordered sequences of events.

However, in real-world processes there usually exist multiple potential identifiers. A single case notion leads to a loss of information (i.e.  \textit{deficiency}, \textit{convergence} and \textit{divergence} issues \cite{vanderAalst2022}). Object-centric process mining \cite{ghahfarokhi_ocel_2021} is an emerging paradigm that drops the single case assumption and instead assumes events can be associated to any number of objects (cases) of different types, with the aim to overcome the limitations of traditional approaches and provide a more accurate depiction of the actual process. Compared to the strictly ordered linear structure of process instances resulting from the single case notion, object-centric process instances can be naturally represented as directed graphs \cite{adams_defining_2022}. 

This study introduces an approach for anomaly detection in business process that is natively designed for object-centric event logs. We propose an unsupervised machine learning approach based on Graph Neural Networks (GNNs), and specifically on a graph convolutional autoencoder (GCNAE) \cite{liu_bond_2022,yuan_higher-order_2021} architecture.

The approach is illustrated in Figure \ref{fig:approach}. It first reconstructs the dependencies between events within an object-centric event log as a set of attributed graphs, then it uses the GCNAE to compute the events' anomaly scores. We employ a simple heuristic based on the inter-quartile range (IQR) to automatically set the threshold and label the anomalies without the need of prior knowledge of the contamination rate of the data.

The main contribution of this study is the introduction of a novel unsupervised anomaly detection approach for business processes, leveraging GNNs and the enriched event data structure offered by object-centric process mining. We are not aware of any other studies that employ GNN on object-centric events logs for anomaly detection. Our approach does not rely on prior information about the process model, contamination rate, or a clean training set, making it suitable for real-world applications. 

We evaluated the performance of our approach on two different publicly available\footnote{Event logs and source code are available at: \href{https://github.com/niro-a/DAEiOcBPvGNN}{github.com/niro-a/DAEiOcBPvGNN}} object-centric event logs: one a synthetic dataset and the other  a real-life dataset. We measured the performance of the approach across different metrics by injecting various types of anomalies in the event logs. The evaluation results demonstrate that our approach performs well in regard to events activity type and attributes anomalies. They also showed limitations regarding the detection of anomalies in the temporal order of events, which requires further exploration of GNN architectures that can better detect such anomalies.

\section{Preliminaries and Background}
\label{sect:background}

\subsubsection{Object-centric Process Mining.} 
Traditional approaches to process mining involve `flattened' event logs that are based on the assumptions of a single case notion and of each event to be associated to exactly one case \cite{van_der_aalst_object-centric_2019}. This leads to a gap between the real event data and the event log, and specifically to the issues of \textit{deficiency} (deletion of events), \textit{convergence} (duplication of events) and \textit{divergence} (ordering unrelated events) \cite{van_der_aalst_object-centric_2019}. An object-centric event log \cite{ghahfarokhi_ocel_2021} is a collection of events where each event is associated with one or more objects, which may be of different types. Similarly to traditional event logs, each event is also associated to an activity, timestamp, and additional attributes. 
\begin{definition}[Object-centric Event Log] Let $T$ be the universe of timestamps. An object-centric event log $L=(E, O, OT, A, AV, \pi_{\text{type}}, \pi_{\text{time}}, \pi_{\text{trace}}, \pi_{\text{act}}, \pi_{\text{attr}})$ is a tuple where:
\begin{itemize}
  \item $E$ is a set of events, $O$ is a set of objects, $OT$ is a set of object types, $A$ is a set of activities and $AV$ is a set of attribute values,
  \item $\pi_{\text{type}}$: $O \rightarrow OT$ maps each object to an object type,
  \item $\pi_{\text{time}}$: $E \rightarrow T$ maps each event to a timestamp,
  \item $\pi_{\text{trace}}$: $O \rightarrow E^*$ maps each object to a temporally ordered sequence of events,
  \item $\pi_{\text{act}}$ : $E \rightarrow A$ maps each event to its activity,
  \item $\pi_{\text{attr}}$ : $E \nrightarrow AV$ maps each event onto attributes values.
\end{itemize} 
\end{definition}

By modeling the relationship between events and multiple objects of different types, object-centric event logs exhibit a graph structure \cite{berti_graph-based_2022} and also the traditional concepts of cases (i.e. process instances) and variants have been extended, in the object-centric setting, from sequences to graphs \cite{adams_defining_2022}. 

\begin{definition}[Object-centric Process Instance]
Let $L$ be an object-centric event log. Given all the temporally ordered traces of events $\pi_{\text{trace}}(o_i)=\left\langle e_1, \ldots, e_n\right\rangle$ associated to each object $o_i \in L$, an object-centric process instance $P=(E^{\prime}, D)$ is a directed graph with nodes $E^{\prime}$ (representing events), edges $D$ (representing the events temporal dependencies) for a set of traces joined directly or transitively by one or more common events.
\end{definition}

Given this definition, an object-centric event log can be reconstructed as a set of one or more process instances, where process instances are made by set of traces that are connected by common bridge events. This representation is free of convergence, deficiency, or divergence issues \cite{adams_addressing_2023} and is equivalent to the \textit{connected component process execution} found in \cite{adams_defining_2022}. It represents a generalization of the traditional case concept to object-centric event logs. 

\subsubsection{Problem Statement.} 

While most of the traditional process mining literature has focused on detecting anomalous process instances (i.e. cases), we move our focus to detecting anomalous events since, in the context of object-centric process mining, process instances can be overly complex and large, to the extreme of comprising one single instance for the whole event log \cite{adams_defining_2022}.

Unlike some of the previous machine learning approaches that characterized anomaly detection as a semi-supervised task \cite{junior_anomaly_2020,lahann_lstm-based_2023,nguyen_autoencoders_2019} which assumes the availability of a suitable labeled dataset of normal behavior, we characterize anomaly detection as an unsupervised task. We further impose the requirement for the algorithms to explicitly discriminate the anomalous events from the normal ones.

\begin{definition}[Event Anomaly Detection]
Event anomaly detection is the task of identifying events that deviate significantly from normal behavior in an object-centric event log. Formally, given an object-centric event log $L$, let $E_n \subseteq E$ be the set of normal events and $E_a \subset E$ be the set of anomalous events. The goal of event anomaly detection is to learn a function $f: E \rightarrow \{0, 1\}$ that assigns a binary label to each event $e_i \in E$ indicating whether it is anomalous or not, based only on the knowledge of $L$. 
\end{definition}

\section{Related Work}
\label{sect:relatedwork}

\subsubsection{Anomaly Detection in Business Processes.} The task of anomaly detection involves identifying observations that do not follow a pattern of normal behavior \cite{chandola_anomaly_2009}, or more specifically in the context of business processes and of process mining, of normal process behavior \cite{ko_systematic_2023}. The exact notion of normal behavior and conversely of anomalous behavior are heavily dependent on the application domain \cite{chandola_anomaly_2009} and on the level of analysis of the specific detection approach \cite{ko_systematic_2023}. A possible categorization of process anomalies is between event-level and process instance-level anomalies \cite{ko_systematic_2023}, where the first refers to anomalies in one or more attributes of an event, while the latter to anomalies in the order or attribute dependencies of events belonging to the same process instance. Following the more general taxonomy found in \cite{chandola_anomaly_2009}, \cite{bohmer_anomaly_2017} classifies process anomalies into three categories based on their nature: \textit{point anomalies}, \textit{contextual anomalies}, and \textit{collective anomalies}. Point (or global) anomalies refer to individual observations that are anomalous compared to the rest of the data, while contextual (or local) anomalies are observations that are only anomalous in specific contexts. Collective anomalies are sets of related observations that are anomalous compared to the entire dataset, even if the individual observations may not be anomalies on their own. 

In regard to existing approaches, \cite{adams_precision_2021} have proposed a generalization of the traditional conformance checking concepts of precision and fitness to object-centric process mining, but otherwise approaches to anomaly detection in object-centric event logs are, to the best of our knowledge, currently unexplored.  More research has been done in the context of traditional process mining, which has mostly focused on detecting anomalous cases (i.e. process instances) \cite{ko_systematic_2023}.

Conformance checking based approaches revolve around detecting anomalous behavior in an event log compared to a reference process model or a discovered process model. These approaches can either focus on a control-flow perspective \cite{bezerra_anomaly_2009,bezerra_algorithms_2013} or consider also additional event attributes \cite{bohmer_multi-perspective_2016,bergami_aligning_2021}.

Distance-based approaches group traces based on distance measures, either via clustering \cite{junior_anomaly_2020} or classification \cite{tavares_analysis_2020} algorithms, while reconstruction-based approaches employ autoencoder neural networks to compute the reconstruction errors for the case encodings, which are used as anomaly scores. Approaches applying standard autoencoders \cite{nolle_analyzing_2018,nguyen_autoencoders_2019} involve representing each case as an ordered vector of event activity types and events attributes, eventually padding shorter cases to the maximum length found in the event log. Similarly, approaches based on recurrent neural networks leverage the sequential nature of the traditional case concept to train autoencoders based on gated-recurrent units (GRUs) \cite{nolle_binet_2022} or long-short term memory (LSTM) neural networks \cite{nguyen_autoencoders_2019,lahann_lstm-based_2023}. \cite{huo_graph_2021} proposed an approach based on graph autoencoders that represented cases as loops and self-loops between activity types.

A common characteristic of the aforementioned approaches is that they require the definition of a threshold to discriminate between anomalous and normal cases \cite{ko_systematic_2023} derived from domain knowledge or some heuristic. 

\subsubsection{Graph Neural Networks.}
GNNs are a class of neural networks designed to operate on graphs. They have been applied successfully to various tasks, including anomaly detection \cite{liu_bond_2022}. In general terms, GNNs are based on learning representations of the nodes of graphs and of their neighborhood (a $k$-hop of connected nodes) via a local function that is invariant to permutation of the neighboring nodes ordering  \cite{velickovic_everything_2023}. \cite{bronstein_geometric_2021} categorizes most GNNs into three classes based on their local function: convolutional, attentional, and generic message-passing. We are not aware of GNNs applications to object-centric process mining, but, in the context of traditional process mining, they have been employed in approaches to process discovery \cite{sommers_process_2021}, predictive process mining \cite{chiorrini_exploiting_2022,harl_explainable_2020,weinzierl_exploring_2022} and anomaly detection \cite{huo_graph_2021}. These approaches have relied on encodings of process instances as ordered sequences of connected nodes, with the exception of \cite{chiorrini_exploiting_2022}, who employed a technique proposed in \cite{diamantini_building_2016} to embed some temporal loops and events parallelism in the process instances, and of \cite{huo_graph_2021} who, as mentioned, encoded process instances as loops and self-loops between activity types (instead of events).

\section{Method}
\label{sect:method}

\subsubsection{Approach Overview.}
\begin{figure*}[b]
    \centering
    \vspace{-30pt}
    \includegraphics[width=1\textwidth]{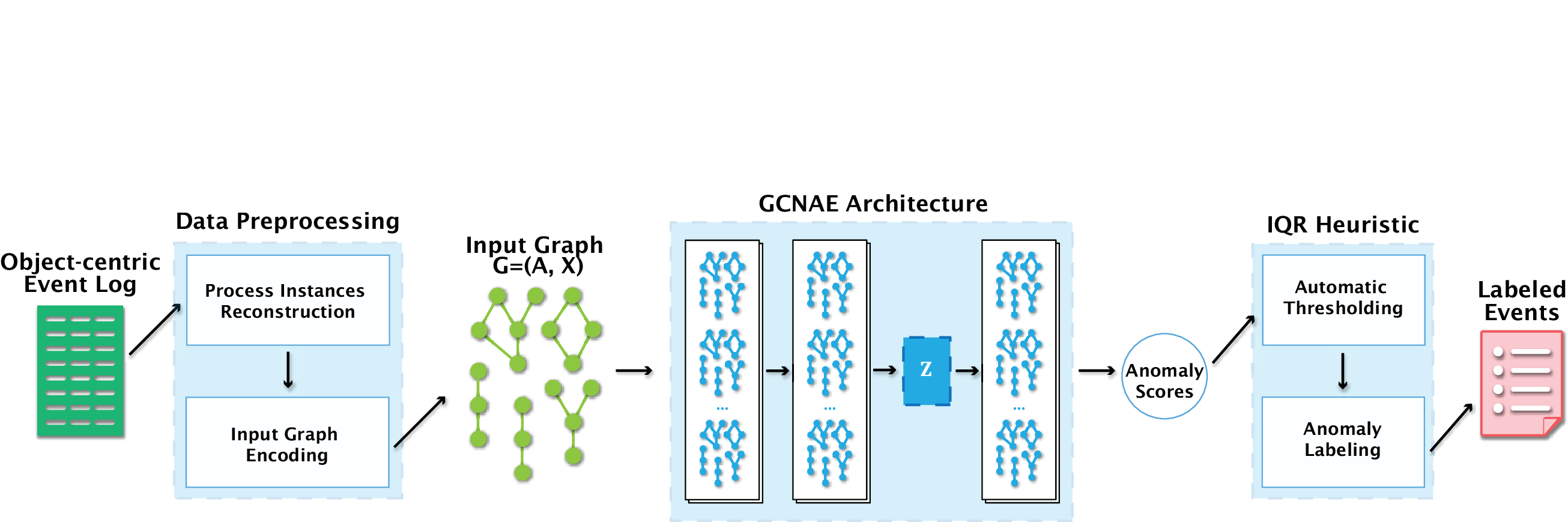}
    \caption{An overview of our proposed approach.}
    \label{fig:approach}
\end{figure*}

Our approach, as illustrated in Figure \ref{fig:approach}, relies on a GCNAE \cite{yuan_higher-order_2021} trained on object-centric event logs containing both normal and anomalous events. We first reconstruct the object-centric process instances from the event logs as a single (disconnected) graph  that serves as input for the GCNAE. The GCNAE is trained to reconstruct the nodes attributes of the input graph. The nodes' reconstruction errors serve as the anomaly scores. We finally apply a simple heuristic based on the IQR to automatically assign a binary label to each event indicating whether it is anomalous or not, without the need to manually set an anomaly score threshold. The following sub-sections explain the different steps in detail.

\subsection{Data Preprocessing}
\label{subsect:method_preprocessing}
\begin{table}[b]
\centering
\caption{An object-centric event log with two object types (A, B).}
\begin{tabular}{cccccccc}
\hline
Event ID & Timestamp       & Obj. Type A & Obj. Type B & Activity & $\text{Attr}_1$ & $\text{Attr}_2$ & \dots \\
\hline
$e_1$        & 01/04/23 09:01 & $a_1$         &               & $\text{act}_1$    & 0.12 & 0.75 & \dots      \\
$e_2$        & 01/04/23 09:07 & $a_2$         & $b_1, b_2$    & $\text{act}_1$    & 0.33 & 0.98 & \dots      \\
$e_3$        & 01/04/23 09:14 &               & $b_1$, $b_2$         & $\text{act}_2$    & 0.24 & 0.39 & \dots      \\
$e_4$        & 01/04/23 09:22 & $a_1, a_3$    & $b_3$         & $\text{act}_3$    & 0.15 & 0.67 & \dots      \\
$e_5$        & 01/04/23 09:37 & $a_2$         &               & $\text{act}_3$    & 0.89 & 0.21 & \dots      \\
$e_6$        & 01/04/23 09:44 & $a_2$         &  $b_2$    & $\text{act}_4$    & 0.58 & 0.46 & \dots      \\
$e_7$        & 01/04/23 10:02 & $a_1, a_3$    &               & $\text{act}_5$    & 0.73 & 0.81 & \dots      \\
$e_8$        & 01/04/23 10:09 &               & $b_3$         & $\text{act}_4$    & 0.42 & 0.34 & \dots      \\
\hline
\end{tabular}
\label{tab:ocel}
\end{table}
\begin{wrapfigure}{R}{0.45\textwidth}
  \centering
   \vspace{-30pt}
  \includegraphics[width=0.4\textwidth]{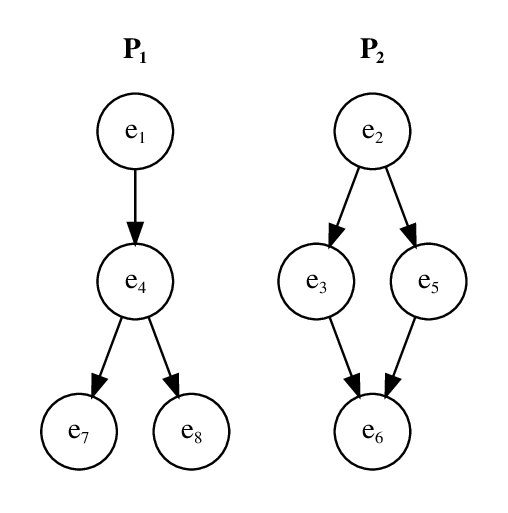}
  %\vspace{-20pt}
  \caption{The reconstructed process instances for the object-centric event log in Table \ref{tab:ocel}.}
  \vspace{-14pt}
  \label{fig:sample-graph}
\end{wrapfigure}

\subsubsection{Process Instances Reconstruction.}
\label{subsubsect:method_prepro_extraction}

We reconstruct the object-centric process instances via the \textit{ocpa} Python library \cite{adams_ocpa_2022} to represent the dependencies between events in the object-centric event log.

As an example, given the simple object-centric event log $L$ in Table \ref{tab:ocel}, the resulting process instances $\mathrm{P_1}$ and $\mathrm{P_2}$ can be visualized in Figure \ref{fig:sample-graph}, where $\mathrm{P_1}$ is composed by the set of traces $\pi_{\text{trace}}(a_1)\cup \pi_{\text{trace}}(a_3) \cup \pi_{\text{trace}}(b_3)=\left\langle e_1, e_4, e_7\right\rangle \cup \left\langle e_4, e_7\right\rangle \cup \left\langle e_4, e_8\right\rangle$ which are bridged by events $e_4$ and $e_7$ and where $\mathrm{P_2}$ is composed by the set of traces $\pi_{\text{trace}}(a_2) \cup \pi_{\text{trace}}(b_1) \cup \pi_{\text{trace}}(b_2)=\left\langle e_2, e_5, e_6\right\rangle \cup \left\langle e_2, e_3\right\rangle \cup \left\langle e_2, e_3, e_6\right\rangle$ which are bridged by events $e_2$, $e_3$ and $e_6$.

\subsubsection{Input Graph Encoding.}
\label{subsubsect:method_prepro_encoding}
We combine the process instances into a single graph $\mathcal{G}$, composed of a set of disconnected subgraphs (i.e., the process instances). This way, we do not need to apply padding or other transformations to the instance graphs. For the example in Figure \ref{fig:sample-graph}, therefore the input graph would correspond to the set of the two subgraphs $P_1$ and $P_2$, $\mathcal{G}_L=\{P_1, P_2\}$. 

\begin{definition}[Adjacency Matrix, $\mathbf{A}$] Given the input graph $\mathcal{G}$ for the event log $L$, where
$\mathcal{G}=(\mathcal{V}, \mathcal{E})$ is a directed graph with nodes $\mathcal{V}=E_L$ and edges $\mathcal{E} \subseteq \mathcal{V} \times \mathcal{V}$, we represent it with an adjacency matrix $\mathbf{A} \in \mathbb{R}^{|\mathcal{V}| \times|\mathcal{V}|}$ where, for an ordered pair of nodes $(u,v)$, $a_{u,v}$ is equal to 1 if $(u, v) \in \mathcal{E}$ and 0 otherwise. 
\end{definition}

\begin{definition}[Feature Matrix, $\mathbf{X}$]
Given an object-centric event log $L$ and its input graph $\mathcal{G}=(\mathcal{V}, \mathcal{E})$, we map to each node $u \in \mathcal{V}$ a feature vector, $\mathbf{x}_u \in \mathbb{R}^k$ corresponding to the event activity type $\pi_{\text{act}}(u)$ and original attributes $\pi_{\text{attr}}(u)$, where categorical features are \textit{one-hot} encoded, and thus $k$ is equal to the sum of unique activity types in $A_L$, unique categorical values in $AV_L$ and numerical attributes in $AV_L$. Then, we stack all the feature vectors into a feature matrix $\mathbf{X} \in \mathbb{R}^{|\mathcal{V}| \times k}$, where $\mathbf{X}=\left[\mathbf{x}_1, \mathbf{x}_2, \ldots, \mathbf{x}_{|\mathcal{V}|}\right]^{\top}$. 
\end{definition}

Accordingly, the input graph $\mathcal{G}$ for the GCNAE is encoded as $\mathcal{G} = (\mathbf{A}, \mathbf{X})$. For the running example of the object-centric event log $L$ in Table \ref{tab:ocel}, the adjacency matrix $\mathbf{A}_L$ and the feature matrix $\mathbf{X}_L$ would look as:

\[
\begin{array}{cc}
\mathbf{A}_L = \begin{bmatrix}
0 & 0 & 0 & 1 & 0 & 0 & 0 & 0 \\
0 & 0 & 1 & 0 & 1 & 0 & 0 & 0 \\
0 & 0 & 0 & 0 & 0 & 1 & 0 & 0 \\
0 & 0 & 0 & 0 & 0 & 0 & 1 & 1 \\
0 & 0 & 0 & 0 & 0 & 1 & 0 & 0 \\
0 & 0 & 0 & 0 & 0 & 0 & 0 & 0 \\
0 & 0 & 0 & 0 & 0 & 0 & 0 & 0 \\
0 & 0 & 0 & 0 & 0 & 0 & 0 & 0 \\
\end{bmatrix}
& 
\mathbf{X}_L = \begin{bmatrix}
1 & 0 & 0 & 0 & 0 & 0.12 & 0.75 & \cdots \\
1 & 0 & 0 & 0 & 0 & 0.33 & 0.98 & \cdots \\
0 & 1 & 0 & 0 & 0 & 0.24 & 0.39 & \cdots \\
0 & 0 & 1 & 0 & 0 & 0.15 & 0.67 & \cdots \\
0 & 0 & 1 & 0 & 0 & 0.89 & 0.21 & \cdots \\
0 & 0 & 0 & 1 & 0 & 0.58 & 0.46 & \cdots \\
0 & 0 & 0 & 0 & 1 & 0.73 & 0.81 & \cdots \\
0 & 0 & 0 & 1 & 0 & 0.42 & 0.34 & \cdots \\
\end{bmatrix}
\end{array}
\]

\subsection{GCNAE Architecture}
\label{subsect:method_architecture}
The GCNAE uses a graph convolutional network (GCN) \cite{kipf_semi-supervised_2017} component both for the the encoder and for the decoder. Graph convolution aggregates the information from neighboring nodes and updates node representations based on their local graph structure.

The encoder component of the GCNAE learns a latent representation of the input graph while the decoder learns to reconstruct the nodes features from this latent representation. The GCNAE is trained by minimizing the reconstruction error between the original nodes features and the reconstructed ones. \\

Adapted from \cite{yuan_higher-order_2021}, our encoder consists of a two-layer GCN:
\begin{equation}
\mathbf{Z} = GCN(\mathbf{X,A}) =  \text{ReLU}(\mathbf{\tilde{A}} \text{ReLU}(\mathbf{\tilde{A}XW_{(0)}})\mathbf{W_{(1)})}
\end{equation}

where the output $\mathbf{Z}$ is a matrix of node embeddings, $\mathbf{X}$ is the feature matrix, $\mathbf{A}$ is the adjacency matrix of the graph. $\text{ReLU}$ is the rectified linear unit activation function with $\text{ReLU(x)}=\text{max(0,x)}$. $\mathbf{\tilde{A}}$ is the symmetrically normalized adjacency matrix with $\mathbf{\tilde{\mathbf{A}}=\mathbf{D}^{-\frac{1}{2}} \mathbf{(A+I)D}^{-\frac{1}{2}}}$ (where $\mathbf{I}$ is the identity matrix and $\mathbf{D}$ is the diagonal degree matrix), used to add self-loops and normalize nodes features aggregation. $\mathbf{W_{(l)}}$ denotes a learnable weight matrix for each $l$-th layer.

The decoder's reconstructed node attributes $\mathbf{\hat{X}}$ are computed with a third GCN layer that maps the node embeddings back to the original input space:

\begin{equation}
\hat{\mathbf{X}}=\text{ReLU}(\mathbf{\tilde{A}ZW_{(2)}})
\end{equation} 

As the loss function to optimize the GCNAE we take the average row-wise mean squared error (MSE) between $\mathbf{X}$ and $\mathbf{\hat{X}}$, while, to account for different shapes in the features encoding, we compute the events anomaly scores by taking the average MSE for each event original feature which are then averaged again.

\subsection{IQR Heuristic: Automatic Thresholding and Anomaly Labeling}
The IQR is is often used to detect outliers and is defined as the difference between the first quartile ($Q_1$) and the third quartile ($Q_3$) of a dataset, ${IQR} = Q_3 - Q_1$. 

To determine a threshold to label anomalies, we compute the $Q_3$ and the IQR for the set of anomaly scores generated by the GCNAE and set the threshold $\tau$ as:
\begin{equation}
\tau = Q_3 + k \cdot \text{IQR} \\
\end{equation} 

Once the threshold is set, we can automatically label the events in the event log as normal or anomalous based on whether their anomaly score is respectively below or above the threshold. In this study, we set $k=1.5$ based on convention, but this value can be adjusted to tailor the sensitivity of the threshold to specific uses.

\section{Experiments}
\label{sect:experiments}

We evaluate our proposed approach on both synthetic and real object-centric event logs, since real event logs can help prove the feasibility of the approach but at the same can also contain unlabeled real-life anomalies which can impact the reliability of the results \cite{nguyen_autoencoders_2019,nolle_binet_2022}. Like in previous work \cite{nguyen_autoencoders_2019,nolle_binet_2022,nolle_analyzing_2018,lahann_lstm-based_2023}, we introduce artificial anomalies into the event logs to simulate various types of irregularities that can occur in real-world processes. Summary statistics for the datasets and the reconstructed process instances can be found in Table \ref{tab:dataset-summary}.

\begin{table*}[b]
\centering
\setlength{\tabcolsep}{7pt}

\caption{Summary statistics for the datasets, the reconstructed process instances and the injected anomalies.}
\resizebox{\textwidth}{!}{%
\begin{tabular}{lccccc}
\hline
\textbf{Dataset} & 
\textbf{\begin{tabular}[c]{@{}c@{}}Original \\ Events\end{tabular}} & 
\textbf{\begin{tabular}[c]{@{}c@{}}Final \\ Events\end{tabular}} & 
\textbf{\begin{tabular}[c]{@{}c@{}}Injected\\ Anomalies\end{tabular}} & 
\textbf{\begin{tabular}[c]{@{}c@{}}Process\\ Instances\end{tabular}} & 
\textbf{\begin{tabular}[c]{@{}c@{}}Events per P. Instance\\ (max, min, avg) \end{tabular}} \\ \hline
\renewcommand{\arraystretch}{1.5}
\begin{tabular}[c]{@{}l@{}}BPIC 2017\end{tabular} & 
393931 & 
407499 & 
40704 & 
31509 & 
(45, 6, 12.9)  \\ \hline
\renewcommand{\arraystretch}{1.5}
\begin{tabular}[c]{@{}l@{}}DS2\end{tabular} & 
22367 &
23137 &
2310 &
83 & 
(1434, 8, 278.8) \\ \hline

\end{tabular}%
}
\label{tab:dataset-summary}
\end{table*}

We use an object-centric version of the \textit{BPIC 2017} dataset \cite{van_dongen_bpi_2017} as a representative of a real event log. It contains almost 400.000 events for the loan application process of a Dutch financial institution between 2016 and 2017. The event log has two types of objects (the application and the offer) and thirteen attributes.

The second event log is the synthetic \textit{DS2} dataset found in \cite{adams_defining_2022}. It simulates an order management process with an especially high amount of connected objects and variability. The event log has three types of objects (items, orders, and packages) and four attributes.

\subsection{Anomalies Injection}
\label{subsect:dataset_injection}

Before reconstructing the process instances, we directly manipulate the object-centric event logs by introducing three types of anomalies in equal parts, amounting in total to circa 10\% of the final number of events in each event log. 

\textit{Attributes Swap}: these anomalies are created by altering the attributes of an event, introducing inconsistencies when compared to events in the same process instance. Given the candidate event $i$, we select the event $j$  whose attributes deviate the most from the attributes of event $i$ by maximizing the euclidean distance $||x_i - x_j||$ and replace the original attributes of $i$ with those of $j$. 

\textit{Timestamp Shift}: these anomalies are introduced by sampling an existing event in the event log and shifting its timestamp within the time-frame ($\pm 5\%$) of all other events that share a common object with the candidate event. This leads to discrepancies in the temporal order of the events in the process instance.

\textit{Random Activities}: similarly to \cite{nolle_binet_2022}, we inject events into the process instances with an activity type that does not come from the original process, while the attributes are sampled from the target process instance.

\subsection{Baselines}
\label{subsect:eval_baselines}
We compare the performance of our graph-based approach to existing approaches for traditional event logs. To be able to do so, we `flatten' the object-centric process instances to temporally ordered sequences composed by the set of events belonging to each process instance. While this flattening strategy still leads to divergence issues, it does not lead to the deletion or duplication of events \cite{adams_addressing_2023}.

As baselines, we implement a standard autoencoder (AE) similar to the one in \cite{nolle_analyzing_2018,nguyen_autoencoders_2019} and a LSTM autoencoder (LSTMAE) similar to the recurrent neural network approaches found in \cite{nolle_binet_2022,nguyen_autoencoders_2019,lahann_lstm-based_2023}. For both, we used the hyperparameters found in previous implementations \cite{nolle_binet_2022,nguyen_autoencoders_2019} in the literature.

\subsection{Results}
\label{subsect:eval_results}

Table \ref{tab:results} shows the experiments results. We ran our experiments five times with different random seeds. To compute the F1 Scores we applied the IQR heuristic to all models, while the other metrics are computed on the raw anomaly scores.

\begin{table}[h!]
\vspace{-1.5pt}
\centering
\setlength{\tabcolsep}{4.4pt}
\caption{Performance comparison of the different models. Results reported as mean $\pm$ standard deviation. Best model in bold.}
\begin{tabular}{llccccc}
\hline
\textbf{Dataset} &
  \multicolumn{1}{l}{\textbf{Model}} &
  \textbf{F1 Score} &
  \textbf{AUC ROC} &
  \multicolumn{1}{c}{\textbf{AUC PR}} &
  \multicolumn{1}{c}{\textbf{Recall @ 10}} \\
  \hline
BPIC 2017 & AE               & 52.4 $\pm$ 0.2         & \textbf{85.6 $\pm$ 0.1} & 43.5 $\pm$ 0.3         & 52.1 $\pm$ 0.2                   \\
          & LSTMAE       & 44.6 $\pm$ 0.3         & 82.9 $\pm$ 0.1 & 34.3 $\pm$ 0.2          & 43.9 $\pm$ 0.2                     \\
          & GCNAE               & \textbf{61.2 $\pm$ 0.2} & 82.4 $\pm$ 0.1 & \textbf{60.3 $\pm$ 0.2} & \textbf{64.8 $\pm$ 0.3}  \\ 
\hline
DS2       & AE               & OOM          & OOM  & OOM         & OOM                    \\
          & LSTMAE       & 31.1 $\pm$ 0.6         & 68.4 $\pm$ 0.4 & 23.5 $\pm$ 0.8         & 30.6 $\pm$ 0.7         &         \\
          & GCNAE               & \textbf{59.2        $\pm$ 0.3} & \textbf{82.5 $\pm$ 0.3} & \textbf{64.9 $\pm$ 0.4} & \textbf{66.6 $\pm$ 0.1}  \\ 
\hline
\end{tabular}
\label{tab:results}
\end{table}

The GCNAE generally outperforms the baselines, which could be attributed to the inherent ability of GNNs to reason over graphs and account for dependencies between events. We also note how the AE goes out-of-memory on the \textit{DS2} dataset since the high number of events in the process instances of this dataset result in long vector encodings.

Furthermore, in Table \ref{tab:recall} we present the $Recall@10$ for each type of anomaly and model. Since in this specific instance of the $Recall@k$ metric we set $k$ equal to the amount of ground truth anomalies in the datasets, it can be used to measure how well the models rank the different anomalous events over the normal ones.

\begin{table}[h]
\centering
\setlength{\tabcolsep}{7pt}
\caption{Recall@10 for the different type of anomalies. Results reported as mean $\pm$ standard deviation.  Best model in bold.}
\begin{tabular}{llcccccc}
\hline
\textbf{Dataset} &
  \multicolumn{1}{l}{\textbf{Model}} &
  \textbf{Attr. Swap} &
  \textbf{Timestamp Shift} &
  \textbf{Random Act.} \\
  \hline
BPIC 2017 & AE     & 89.1 $\pm$ 0.5  & \textbf{25.5 $\pm$ 0.1}  & 41.8 $\pm$ 0.4   \\
          & LSTMAE & 81.8 $\pm$ 0.4  & 18.8 $\pm$ 0.2  & 30.9 $\pm$ 0.6   \\
          & GCNAE  & \textbf{96.9 $\pm$ 0.4}  & 4.0 $\pm$ 0.1   & \textbf{93.4 $\pm$ 0.6}  \\ 
\hline
DS2       & AE     & OOM               & OOM                & OOM                    \\
          & LSTMAE & 76.3 $\pm$ 2.8  & \textbf{7.4 $\pm$ 0.6}   & 8.0 $\pm$ 1.3   \\
          & GCNAE  & \textbf{96.4 $\pm$ 0.1}  & 3.6 $\pm$ 0.4   & \textbf{100.0 $\pm$ 0.0}  \\ 
\hline
\end{tabular}
\label{tab:recall}
\end{table}

The GCNAE struggles detecting the \textit{Timestamp Shift} anomalies. A possible explanation is that GCNs learn nodes representations by aggregating local neighborhood information. Shifting an event within a process instance time-frame might create very subtle changes in the event neighborhood, which therefore would limit the performance of the GCNAE for this specific anomaly. 

\section{Conclusion}
\label{sect:conclusion}

Detecting anomalies is important for identifying inefficiencies, errors, or fraud in business processes. Object-centric event logs provide benefits over traditional event log representations. We have presented a novel approach for anomaly detection in business processes that can leverage the joint capabilities of GNNs and object-centric process mining. Our approach has displayed promising performance, while also not requiring a process model, knowledge of the contamination rate, or the availability of a clean training set.

Future work could explore the use of GNNs architectures that are better suited to learn the temporal and structural dependencies of business process instances and address the limitations displayed by the GCNAE. Future work could also extend the focus to additional anomaly types and object-centric perspectives, for example moving the level of analysis to anomalous process instances and sub-process instances, or investigating anomalies arising in the relationships between and within multiple objects.

% \newpage

\bibliographystyle{splncs04}
\bibliography{references}

\end{document}